\documentclass[10pt,prb,twocolumn,superscriptaddress,amsmath,amssymb,floatfix,preprintnumbers]{revtex4-2}
\usepackage[utf8]{inputenc}
\usepackage[T1]{fontenc}
\usepackage{amssymb,amsmath,bm}
\usepackage[pdftex]{graphicx}
\usepackage{color,soul}
\usepackage{tikz,xcolor,hyperref}
\usepackage{mathtools}
\usepackage{amsmath}
\usepackage{dcolumn}
\usepackage{bm}
\usepackage{xcolor}
\usepackage{float}
\usepackage[mathlines]{lineno}
\usepackage{cleveref}
\usepackage{listings}
\usepackage{lipsum}
\usepackage{bbold}
\hypersetup{colorlinks=true,linkcolor=blue,citecolor=blue}
\usepackage{orcidlink}
\usepackage{braket}
\newcommand{\bfrac}[2]{\genfrac{}{}{0pt}{}{#1}{#2}}
\hypersetup{colorlinks=true,linkcolor=blue,citecolor=blue}

\begin{document}

\title{Valley-selective confinement of excitons in transition metal dichalcogenides with inhomogeneous magnetic fields}

\author{A. J. Chaves\,\orcidlink{0000-0003-1381-8568}}
\email{andrejck@ita.br}
\affiliation{Department of Physics, Aeronautics Institute of Technology, 12228-900, São José dos Campos, SP, Brazil}
\affiliation{POLIMA---Center for Polariton-driven Light--Matter Interactions, University of Southern Denmark, Campusvej 55, DK-5230 Odense M, Denmark}

\author{D. R. da Costa \orcidlink{0000-0002-1335-9552}}
\email{diego\_rabelo@fisica.ufc.br}
\affiliation{Departamento de F\'isica, Universidade Federal do Cear\'a, Campus do Pici, 60455-900 Fortaleza, Cear\'a, Brazil}
\affiliation{Department of Physics, University of Antwerp, Groenenborgerlaan 171, B-2020 Antwerp, Belgium}

\author{F. M. Peeters\orcidlink{0000-0003-3507-8951}}
\email{francois.peeters@uantwerpen.be}
\affiliation{Departamento de F\'isica, Universidade Federal do Cear\'a, Campus do Pici, 60455-900 Fortaleza, Cear\'a, Brazil}
\affiliation{Department of Physics, University of Antwerp, Groenenborgerlaan 171, B-2020 Antwerp, Belgium}

\author{Nuno~M.~R.~Peres\,\orcidlink{0000-0002-7928-8005}}
\email{peres@fisica.uminho.pt}
\affiliation{POLIMA---Center for Polariton-driven Light--Matter Interactions, University of Southern Denmark, Campusvej 55, DK-5230 Odense M, Denmark}
\affiliation{Centro de F\'{\i}sica (CF-UM-UP) and Departamento de F\'{\i}sica, Universidade do Minho, P-4710-057 Braga, Portugal}

\begin{abstract}
Magnetized ferromagnetic disks or wires support strong inhomogeneous fields in their borders. Such magnetic fields create an effective potential, due to Zeeman and diamagnetic contributions, that can localize charge carriers. For the case of two-dimensional transition metal dichalcogenides, this potential can valley-localize excitons due to the Zeeman term, which breaks the valley symmetry. We show that the diamagnetic term is negligible when compared to the Zeeman term for monolayers of transition metal dichalcogenides. The latter is responsible for trapping excitons near the magnetized structure border with valley-dependent characteristics, in which, for one of the valleys, the exciton is confined inside the disk, while for the other, it is outside. This spatial valley separation of exciton can be probed by circularly polarized light, and moreover, we show that the inhomogeneous magnetic field magnitude, the dielectric environment, and the magnetized structure parameters can tailor the spatial separation of the exciton wavefunctions. 
\end{abstract}

\maketitle

\section{Introduction}\label{sec:introduction}

The trapping of quantum particles is present in different nanosystems, such as quantum dots, rings, wires, and wells, to compose the essential building blocks for several future technological applications. \cite{harrison2016quantum, chakraborty1999quantum} A diverse range of trapping approaches for charged carriers and excitons by defects \cite{song1993investigation}, external electrostatic gates \cite{schinner2012single, PhysRevApplied.19.044095}, and external homogeneously applied magnetic field \cite{manninen2001magnetic, lee2004magnetic} have been demonstrated in the 90s and 2000s in two-dimensional electron gas (2DEG) systems. Among other various methods, the possibility of trapping and guiding charged particles by means of nonhomogeneous magnetic fields has attracted considerable interest. \cite{nogaret1997observation, Nogaret_2010, christianen1998magnetic, pulizzi2000two, geim1997ballistic, fleurov2002spin, reijniers1998hybrid, ibrahim1998diffusive, uzur2004probing, Freire2000} For example, A. Nogaret \textit{et al.} \cite{nogaret1997observation, Nogaret_2010} have reported clear evidence for the channeling of 2D electrons in open orbits along the lines of zero magnetic field, obtained by an array of nickel stripes fabricated by electron-beam lithography directly on the surface of the 2DEG heterostructure. For trapping excitons, Refs.~\cite{christianen1998magnetic, pulizzi2000two, geim1997ballistic} showed the 2D exciton control by using a thin magnetized stripe positioned on top of quantum wells. In general, they demonstrated that a strong magnetic field gradient can force excitons to regions of low-field gradient working as magnetic traps for excitons. In this context, one of the examples that motivated our current work was that proposed by J. A. K. Freire \textit{et al.} \cite{Freire2000} that used a magnetized disk to trap excitons at its borders.

\begin{figure*}[!t]
\includegraphics[width=0.8\linewidth]{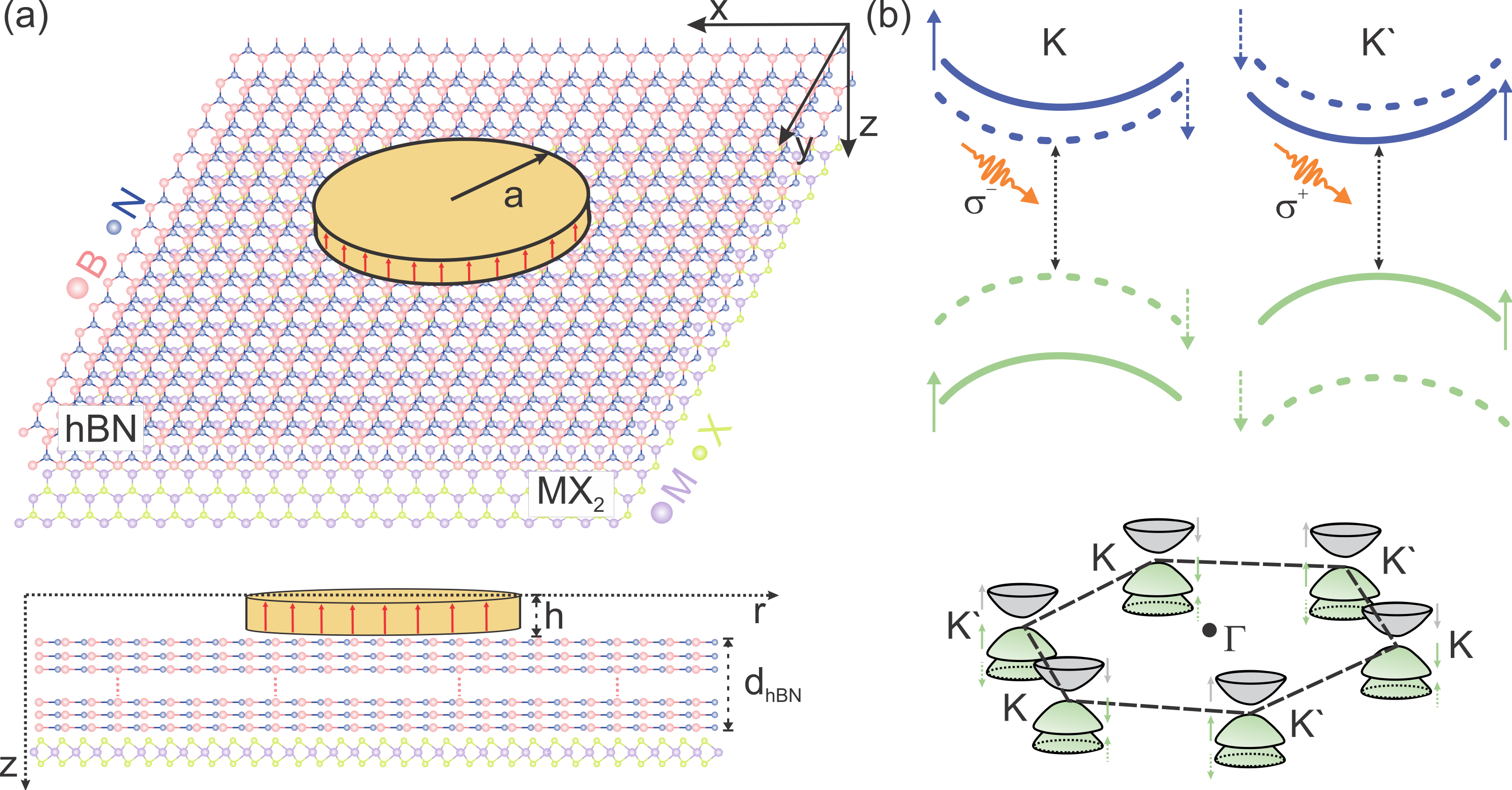}
\caption{\textcolor{blue}{(Color online)} (a) Sketch of the exciton trapping mechanisms composed by a MX$_2$ sheet covered by an hBN slab where it is deposited on a circular magnetized disk with radius $a$ that induces an inhomogeneous magnetic field in the MX$_2$ layer. The distance between the magnetized disk and the TMD layer is $z = h + d_{hBN}$, where $h$ is the disk height and $d_{hBN}$ is the thickness of the hBN slab formed by $n$ hBN layers. Once the confined excitons are localized at $z$, the hBN slab will modulate the environmental dielectric screening on atomic length scales. (b) Valley- and spin-dependent optical transition selection rules in monolayer TMD, illustrating that the direct bandgap optical transition at $K$ ($K^\prime$) valley couples only to $\sigma^+$ ($\sigma^-$) circularly polarized light. The bottom illustration in (b) presents the two non-equivalent gapped Dirac-like cones and the corresponding valley-dependent spin orientations.}
\label{Fig1}
\end{figure*}

From the large family of lamellar materials, a potential candidate for applications in optoelectronics \cite{rasmita2021opto, mak2016photonics, tian2016optoelectronic, mueller2018exciton}, spintronics \cite{ahn20202d, feng2017prospects, xu2014spin}, and valleytronics \cite{schaibley2016valleytronics, vitale2018valleytronics, rasmita2021opto, xiao2012coupled, braganca2019magnetic, liu2019valleytronics} is the group VI transition metal dichalcogenides (TMDs), usually abbreviated by the chemical formula of MX$_2$, where $M$ and $X$ denote a transition metal (e.g., $Mo$ and $W$) and a chalcogen (e.g., $S$ and $Se$), respectively [see Fig.~\ref{Fig1}(a)]. Both the bulk and monolayer TMDs are semiconductors, presenting a thickness-dependent band structure tunability, with the bandgap increasing monotonically with decreasing the number of layers and, most interestingly, exhibiting cross-over characteristics that change from an indirect (bulk) to direct (monolayer) bandgap material with gaps located at the $K$ and the $K^\prime$ points [see bottom sketch in Fig.~\ref{Fig1}(b)]. \cite{splendiani2010emerging} Moreover, TMDs' band structures resemble that of massive gapped graphene with a strong spin-orbit coupling that splits the spin degeneracy of the bands and locks their spin and valley pseudospin degrees of freedom. \cite{xiao2012coupled, xu2014spin} Furthermore, its unique band structure has non-null out-of-plane orbital magnetic moments of opposite signs at the different valleys, which, in turn, give rise to a valley-dependent optical selection rule, absorbing left ($\sigma_+$) or right ($\sigma_-$) circularly polarized light by $K$ or $K^\prime$ valleys [see top sketch in Fig.~\ref{Fig1}(b)]. \cite{cao2012valley, mak2012control, zeng2012valley, sallen2012robust} Due to its direct-gap semiconductor nature, reduced dimensionality, and reduced dielectric screening of Coulomb interactions between charge carriers, one of the key features of the TMDs' family is that they host strong excitonic effects. \cite{he2014tightly, berkelbach2013theory, ye2014probing, ugeda2014giant, cavalcante2018stark, tenorio2023tunable, mueller2018exciton} Experimental observations reported large binding energies and small Bohr radius for the excitonic complexes formed in monolayer TMDs as compared with conventional semiconductors. \cite{ugeda2014giant, zhang2014direct}.

Motivated by the strong optical responses in TMDs that are dominated by excitons and other excitonic complexes, many different approaches have been reported in the literature seeking ways to control and manipulate such electron-hole-bound states. \cite{malic2023exciton, rosati2021dark, niehues2018strain, tedeschi2019controlled, wang2018electrical} Owing to that, trapping methods for excitons involving strain \cite{bensmann2022nanoimprint, kern2016nanoscale, palacios2017large, branny2017deterministic, long2020exciton}, point defects \cite{wang2022intensive, wu2022revealing, chen2017experimental}, spatially finite structures \cite{qu2017tunable, PhysRevB.96.035122}, type-II band alignment in TMD hetero-bilayer \cite{li2020dipolar}, different dielectric environments \cite{raja2017coulomb}, and moiré traps in twist TMD homo- and hetero-bilayers \cite{malic2023exciton, seyler2019signatures} have been proposed. For instance, Y. Wang \textit{et al.} \cite{wang2022intensive} have reported defect-related trap-bound exciton states in suspended WS$_2$ monolayers at cryogenic temperatures, using steady-state and time-resolved photoluminescence spectroscopy. W. Li \textit{et al.} \cite{li2020dipolar} and K. L. Seyler \textit{et al.} \cite{seyler2019signatures} have reported experimental evidence of interlayer valley excitons trapped in a smooth moiré potential with inherited valley-contrasting physics in MoSe$_2$/WSe$_2$ hetero-bilayers. J. Bensmann \textit{et al.} \cite{bensmann2022nanoimprint} have experimentally created an inhomogeneous strain profile in a monolayer TMD on a micrometer scale by a nanoimprint process and found that in regions close to the imprint the exciton energy is red-shifted, where such lower exciton energy corresponds to larger tensile strain. C. Palacios-Berraquero \textit{et al.} \cite{palacios2017large} have experimentally created deterministic arrays of hundreds of quantum emitters in WSe$_2$ and WS$_2$ monolayers, achieved by depositing monolayers onto silica substrates nanopatterned with arrays of nanopillars, and observed that such nanopillars create localized deformations in the TMD material resulting in the quantum confinement of excitons. Similarly, A. Branny \textit{et al.} \cite{branny2017deterministic} have experimentally employed nanoscale strain engineering to achieve a 2D lattice of quantum emitters in TMDs, observing that optically created excitons efficiently funnel to an individual strain-tuned localized exciton trap at the nanopillar center resulting in a single highly efficient quantum emitter.

Combining the mentioned physical ingredients of magnetic-induced quantum confinement widely investigated in conventional semiconductors, the strong excitonic effects, and the spin- and valley-dependent properties of the monolayers' TMDs, here, differently from the previously reported proposals, we show that a mechanism based on a magnetic field gradient induced by a circularly symmetric magnetized disk, which experimentally can be created by the deposition of the disk on top of a hexagonal boron nitride (hBN) slab as illustrated in Fig.~\ref{Fig1}(a), can trap monolayer TMD excitons. To our knowledge, no theoretical study based on an inhomogeneous magnetic field profile has been discussed as an appropriate manner to confine excitons in monolayer TMDs and, moreover, featuring valley- and spin-dependent spatial localization of the exciton energies and wavefunctions outside or inside the disk. To this end, we organized the paper as follows. In Sec.~\ref{sec.magnetic.field}, we derived the vector potential associated with the circularly symmetric magnetized disk [Fig.~\ref{Fig1}(a)] and showed the dependence of the inhomogeneous magnetic field profile on the distance between the TMD layer and the magnetic disk. In Sec.~\ref{sec.excitons}, we presented the effective Hamiltonian for the relative (Sec.~\ref{sub.relative}) and center of mass (Sec.~\ref{sub.center}) motions based on the effective mass and Born-Oppenheimer approximations. Results for the trapping exciton energies and wavefunctions for different disk magnetization, the exciton binding energies and average radius for different dielectric environments controlled by the thickness of the hBN slab, and the trapping potential for different disk thicknesses were discussed. Our final remarks are addressed in Sec.~\ref{sec.discussions}. 

\section{Magnetic field of a disk}\label{sec.magnetic.field}

Before we investigate the exciton confinement due to a magnetized disk, in this section, we shall calculate the magnetic field created by a magnetized and finite circular disk of radius $a$ and height $h$, such as depicted in Fig.~\ref{Fig1}, at a point in the space with an in-plane distance $r$ and out-of-plane distance $z$ from the center of the disk. We consider the magnetostatic regime and choose the Coulomb gauge constraint, i.e., the vector potential $\mathbf{A}$ obeys the relation $\boldsymbol{\nabla}\cdot\mathbf{A}=0,$ therefore, from Ampère's law, one gets
\begin{equation}\label{eq.ampere}
    \nabla^2\mathbf{A}(\mathbf{r})=-\mu_0\mathbf{j}(\mathbf{r}),
\end{equation}
with $\mathbf{j}(\mathbf{r})$ being the total current at the system. From the Laplacian Green's function, $-(4\pi|\mathbf{r}-\mathbf{r}^\prime |)^{-1}$, the vector potential solution of Eq.~\eqref{eq.ampere} can be written as
\begin{equation}
    \mathbf{A}(\mathbf{r})=\frac{\mu_0}{4\pi} \int d^3\mathbf{r}^\prime \frac{\mathbf{j}(\mathbf{r}^\prime)}{|\mathbf{r}-\mathbf{r}^\prime|}. \label{eq:vector_potential}
\end{equation}

As the $\mathbf{j}(\mathbf{r})$ is the total current, one can decompose it in free and bound currents' contributions, such as $\mathbf{j}(\mathbf{r})=\mathbf{j}_\mathrm{free}(\mathbf{r})+\mathbf{j}_M(\mathbf{r})$, where the term $\mathbf{j}_M(\mathbf{r})$ corresponds to the disk magnetization and includes both spin and orbital magnetism. Let $\mathbf{M}(\mathbf{r})$ be the total magnetization, the magnetization bulk current $\mathbf{j}_M(\mathbf{r})$ is
\begin{equation}    \mathbf{j}_M(\mathbf{r})=\boldsymbol{\nabla}\times\mathbf{M}(\mathbf{r}).
\end{equation}
Neglecting contributions from free currents, the vector potential \eqref{eq:vector_potential} reads
\begin{equation}
    \mathbf{A}(\mathbf{r})= \frac{\mu_0}{4\pi} \int d^3\mathbf{r}^\prime \frac{\boldsymbol{\nabla}^\prime\times\mathbf{M}(\mathbf{r}^\prime)}{|\mathbf{r}-\mathbf{r}^\prime|}. \label{eq:Afield}
\end{equation}
Due to the finiteness of the disk, a surface magnetization current $\mathbf{K}(\mathbf{r})=\mathbf{M}(\mathbf{r})\times\mathbf{\hat{n}}(\mathbf{r})$ appears at the disk boundary, whose normal vector is $\mathbf{\hat{n}}$. Considering a constant magnetization inside the disk, therefore solely the surface magnetization contribution survives, resulting in
\begin{equation}
        \mathbf{A}(\mathbf{r})= \frac{\mu_0}{4\pi} \int_S d^2\mathbf{r}^\prime \frac{\mathbf{M}(\mathbf{r}^\prime)\times\mathbf{n}(\mathbf{r}^\prime)}{|\mathbf{r}-\mathbf{r}^\prime|},
\end{equation}
where $\int_S$ denotes the surface integral. By assuming that the disk axis is parallel to the total magnetization $\mathbf{M}$ and taking it constant and along the $z$ direction, such as $\mathbf{M}={\cal M} \mathbf{u}_z$, then the vector product $\mathbf{M}\times\mathbf{\hat{n}}$ is finite only in the lateral surface of the disk. Consequently, only the lateral surface's contribution to the integration in Eq.~\eqref{eq:Afield} will be non-null. Thus, $\mathbf{\hat{n}}(\mathbf{r})=\mathbf{\hat{u}}_r$ points towards the radial direction, and consequently $\mathbf{\hat{u}}_z\times\mathbf{\hat{u}}_r=\mathbf{\hat{u}}_\theta$ and the vector potential is orientated along $\mathbf{\hat{u}}_\theta$, i.e.,
\begin{equation}
    \mathbf{A}(\mathbf{r})=A(r,z)\mathbf{\hat{u}}_\theta,
\end{equation}
where
\begin{align}\label{eq.Arz.before}
    A(r,z)\hspace{-0.075cm}=\hspace{-0.075cm}\frac{\mu_0 a {\cal M}}{4\pi}\hspace{-0.2cm}\int_0^h \hspace{-0.25cm} dz^\prime \hspace{-0.225cm}\int_0^{2\pi} \hspace{-0.4cm}  \frac{\cos\theta^\prime d\theta^\prime}{\sqrt{r^2 \hspace{-0.075cm}+\hspace{-0.075cm} a^2 \hspace{-0.075cm}-\hspace{-0.075cm} 2ra\cos\theta^\prime \hspace{-0.075cm}+\hspace{-0.075cm} (z^\prime \hspace{-0.075cm}-\hspace{-0.075cm} z)^2}}.
\end{align}
\begin{figure}[!t]
\includegraphics[width=\linewidth]{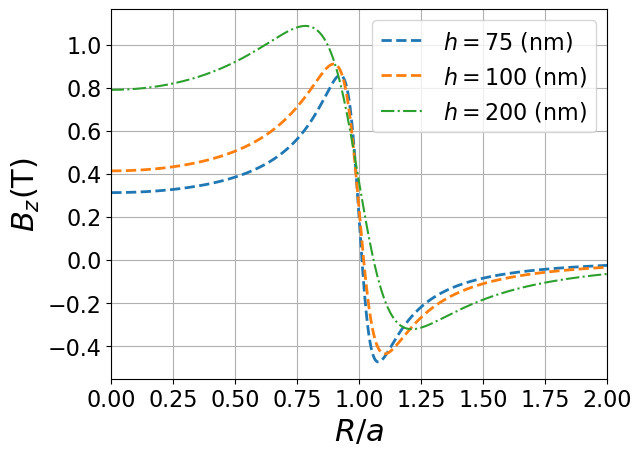}
\caption{\textcolor{blue}{(Color online)} Spatial dependence of the non-homogeneous magnetic field profile $B_z$ [Eq.~\eqref{eq:Bfield.final}] induced by the magnetized disk for different disk thickness $h$, with radius $a=1\,\mu$m and $z=h+d_\mathrm{hBN}$ with $d_\mathrm{hBN}=1.0$ nm, corresponding to three hBN layers, for magnetization of $\mu_0 {\cal M}=1.33$ T and null external magnetic field.}
\label{Fig2}
\end{figure}
By performing the integration in $z^\prime$, one obtains
\begin{align}\label{eq.Arz}
    A(r,z)\hspace{-0.075cm}=\hspace{-0.075cm}\frac{\mu_0 a {\cal M}}{4\pi} \hspace{-0.2cm} \int_0^{2\pi} \hspace{-0.3cm}  d\sin\theta^\prime \ln \hspace{-0.075cm}\left[ \hspace{-0.075cm}\frac{\sqrt{(h \hspace{-0.075cm}-\hspace{-0.075cm} z)^2 \hspace{-0.075cm}+\hspace{-0.075cm} x^2} \hspace{-0.075cm}+ \hspace{-0.075cm} h \hspace{-0.075cm}-\hspace{-0.075cm} z}{\sqrt{z^2 \hspace{-0.075cm}+\hspace{-0.075cm} x^2} \hspace{-0.075cm}-\hspace{-0.075cm} z} \hspace{-0.075cm}\right],     
\end{align}
where $x=\sqrt{r^2+a^2-2ra\cos\theta^\prime}$. By taking the limit $h\rightarrow 0$, it allows us to write Eq.~\eqref{eq.Arz.before} as
\begin{equation}
    A(r,z)\hspace{-0.075cm}=\hspace{-0.075cm}\frac{\mu_0ha{\cal M}}{4\pi} \hspace{-0.2cm} \int_0^{2\pi} \hspace{-0.3cm} \frac{\cos\theta^\prime d\theta^\prime}{\sqrt{r^2 \hspace{-0.075cm}+\hspace{-0.075cm} a^2 \hspace{-0.075cm}+\hspace{-0.075cm} z^2 \hspace{-0.075cm}-\hspace{-0.075cm} 2ra\cos\theta^\prime}},
\end{equation}
and using the change of variables $\theta^\prime=2\alpha$ and manipulating the above expression, one obtains
\begin{align}\label{eq.A.after.manipulations}
    \hspace{-0.2cm} A(r,z) &=\frac{\mu_0ha{\cal M}}{2\pi} \hspace{-0.15cm} \int_0^\pi \hspace{-0.15cm} d\alpha \frac{1 \hspace{-0.075cm}- \hspace{-0.075cm} 2\sin^2\alpha}{\sqrt{1 \hspace{-0.075cm}+\hspace{-0.075cm} 4ra\sin^2\alpha}} \nonumber\\ &= \frac{\mu_0ha{\cal M}}{\pi\sqrt{(r \hspace{-0.075cm}-\hspace{-0.075cm} a)^2 \hspace{-0.075cm}+\hspace{-0.075cm} z^2}} \hspace{-0.15cm} \int_0^{\frac{\pi}{2}} \hspace{-0.15cm} d\alpha \frac{1 \hspace{-0.075cm}-\hspace{-0.075cm} 2\sin^2\alpha}{\sqrt{1 \hspace{-0.075cm}+\hspace{-0.075cm} k^2\sin^2\alpha}},
\end{align}
where $k^2=4ra/[(r-a)^2+z^2]$. Finally, one can express the integral \eqref{eq.A.after.manipulations} in terms of complete elliptic functions of the first ($K$) and second ($E$) types as
\begin{align}
    \hspace{-0.2cm} A(r,z) &= \frac{\mu_0ha{\cal M}}{\pi\sqrt{ra}} \hspace{-0.075cm}\left\{ \hspace{-0.075cm}\frac{k}{\sqrt{k^2+1}} K \hspace{-0.075cm}\left( \hspace{-0.075cm}\frac{k^2}{k^2+1} \hspace{-0.075cm}\right)\nonumber\right. \\ & \left. \hspace{2cm} +\frac{2}{k}\left[K(-k^2) \hspace{-0.075cm}-\hspace{-0.075cm} E(-k^2)\right]\hspace{-0.075cm} \right\}. \label{eq:Afield_c}
\end{align}

Once the magnetic field induced by the magnetized disk that will confine the charge carriers is along the $z$ direction, thus at the end of the day, only its $z$ component is relevant to be calculated, given by
\begin{equation}
    B_z(r)=\frac{1}{r}\partial_r\left[ rA(r,z) \right], \label{eq:Bfield}
\end{equation}
with the analytical expression of this derivative given in Appendix \ref{app:Bz}.

In Fig.~\ref{Fig2}, we show the magnetic field profile $B_z(r)$ for a magnetized disk with radius $a=1.0\,\mu$m and magnetization of $\mu_0{\cal M}=1.33$ T for different values of the disk height $h$, considering that the 2D material is placed at a distance $z=h+d_\mathrm{hBN}$ from the TMD layer, where $d_\mathrm{hBN}=1.0$ nm corresponds to three layers of hBN, commonly used to encapsulate 2D materials \cite{Zhang2017}. The parameters of the disk and magnetization were chosen to be compatible with feasible experimental magnetic disk values, as in Ref.~\cite{Uzur2004}. As the magnetized disk's thickness increases, the magnetic field profile becomes less sharp, and its magnitude slowly increases for greater thicknesses. 

\section{Excitons in a magnetic field}\label{sec.excitons}

In this section, we will obtain the effective Hamiltonian for both the center of mass motion and relative motion of an exciton in a 2D semiconductor material subjected to an external inhomogeneous magnetic field within the effective mass and Born-Oppenheimer approximations. For that, we start with the Hamiltonian for the electron-hole pair inside a magnetic field described by the vector potential $\mathbf{A}(\mathbf{r})$
\begin{equation}\label{eq.H.initial}
H=\frac{\left[\mathbf{p}_e +e\mathbf{A}(\mathbf{r}_e)\right]^2}{2m_e}
+\frac{\left[\mathbf{p}_e-e\mathbf{A}(\mathbf{r}_h)\right]^2 }{2m_h}
+V(\mathbf{r}_e-\mathbf{r}_h),
\end{equation}
where $m_e$ and $m_h$ are the electron and hole effective masses, respectively, $\mathbf{r}_{e(h)}$ the electron (hole) vector position, $\mathbf{p}_{e(h)}$ the electron(hole)-momentum, $e>0$ is the fundamental electron charge, and $V(\mathbf{r})$ is the electron-hole interaction, assumed to be the Rytova-Keldysh potential \cite{Cudazzo2011}. Now, introducing the relative $\mathbf{r}$ and center of mass $\mathbf{R}$ coordinates, such as
\begin{subequations}
    \begin{eqnarray}
        \mathbf{r}&=&\mathbf{r}_e-\mathbf{r}_h,\\
        \mathbf{R}&=&\frac{m_e\mathbf{r}_e+m_h\mathbf{r}_h}{M},
    \end{eqnarray}
\end{subequations}
where $M=m_e+m_h$ is the total mass of the exciton, and the relative $\mathbf{p}$ and center of mass $\mathbf{P}$ momenta, given by
\begin{subequations}
    \begin{eqnarray}
        \mathbf{p}&=&\frac{m_h\mathbf{p}_e-m_e\mathbf{p}_h}{M},\\
        \mathbf{P}&=&\mathbf{p}_e+\mathbf{p}_h,
    \end{eqnarray}
\end{subequations}
obeying the commutator relations $\left[\mathbf{r}_i,\mathbf{p}_j\right]=i\hbar\delta_{ij}$, $\left[\mathbf{r}_i,\mathbf{r}_j\right]=\left[\mathbf{p}_i,\mathbf{p}_j\right]=0$, $\left[\mathbf{R}_i,\mathbf{P}_j\right]=i\hbar\delta_{ij}$, and $\left[\mathbf{R}_i,\mathbf{R}_j\right]=\left[\mathbf{P}_i,\mathbf{P}_j\right]=0$, one can rewrite the Hamiltonian \eqref{eq.H.initial} in terms of the relative and center of mass motions. By using the Coulomb gauge, one has for the kinetic energy term
\begin{flalign}
\frac{\left[\mathbf{p}_{\bfrac{e}{h}}\pm  e\mathbf{A}( \mathbf{r}_{\bfrac{e}{h}} )\right]^2}{2m_{\bfrac{e}{h}}} = \frac{\mathbf{p}^2_{\bfrac{e}{h}} \pm  2e\mathbf{A}(\mathbf{r}_{\bfrac{e}{h}})\cdot\mathbf{p}_{\bfrac{e}{h}}+e^2\mathbf{A}^2(\mathbf{r}_{\bfrac{e}{h}})}{2m_{\bfrac{e}{h}}},
\end{flalign}
and performing the coordinate transformation for the center of mass and relative coordinates, one gets
\begin{align}\label{eq.kinetic}
 &\frac{\left[\mathbf{p}_{\bfrac{e}{h}} \pm e\mathbf{A}(\mathbf{r}_{\bfrac{e}{h}} )\right]^2}{2m_{\bfrac{e}{h}} } = \frac{\mathbf{p}^2}{2m_{\bfrac{e}{h}}} + \frac{m_{\bfrac{e}{h}}}{2M^2}\mathbf{P}^2  \pm \frac{1}{M}\mathbf{p}\cdot\mathbf{P} \nonumber\\ 
 & \hspace{0.65cm} + \frac{e^2\mathbf{A}^2(\mathbf{R},\mathbf{r})}{2m_{\bfrac{e}{h}}} \pm \frac{e}{m_{\bfrac{e}{h}}}\mathbf{A}(\mathbf{R},\mathbf{r})\cdot \left(\mathbf{p} \pm \frac{m_{\bfrac{e}{h}}}{M}\mathbf{P}\right).  
\end{align}
Replacing Eq.~\eqref{eq.kinetic} into Eq.~\eqref{eq.H.initial}, it reads
\begin{align}
    H &=\frac{\mathbf{p}^2}{2\mu}+\frac{\mathbf{P}^2}{2M}+\frac{e^2\mathbf{A}^2(\mathbf{R},\mathbf{r})}{2\mu}+\frac{2e}{M}\mathbf{A}(\mathbf{R},\mathbf{r})\cdot\mathbf{P} +V(r) \nonumber\\
    &+e\left(\frac{1}{m_e}-\frac{1}{m_h}\right)\mathbf{A}(\mathbf{R},\mathbf{r})\cdot\mathbf{p}, \label{eq:total_H}
\end{align}
with $\mu^{-1}=m_e^{-1}+m_h^{-1}$ being the reduced mass. Due to the small electron-hole mass asymmetry for monolayer TMDs \cite{Kormanyos_2015}, we now will assume that $m_e=m_h=m_rm_0$, where $m_r$ is the effective mass and $m_0$ the electron rest mass; as a consequence, the last term of Eq.~\eqref{eq:total_H} vanishes. 

From now on, one considers that the magnetic field changes slowly in the exciton Bohr radius scale, i.e., for the relative coordinate, the magnetic field is constant. This implies that the vector potential can be written as
\begin{equation}
    \mathbf{A}(\mathbf{R},\mathbf{r})=\frac{1}{2}\mathbf{B}(\mathbf{R})\times\mathbf{r},
\end{equation}
in the sense that $\mathbf{B}(\mathbf{R})=\boldsymbol{\nabla}_\mathbf{r}\times\mathbf{A}(\mathbf{R},\mathbf{r})$. Assuming a perpendicular field $\mathbf{B}(\mathbf{R})=B_z(R)\mathbf{u}_z$, one has
\begin{equation}
\mathbf{A}^2(\mathbf{R},\mathbf{r})=\frac{1}{4}B_z^2(R)r^2,
\end{equation}
and 
\begin{equation}
\mathbf{A}(\mathbf{R},\mathbf{r})\cdot\mathbf{P}=\frac{1}{2}\left[\mathbf{B}(R)\times\mathbf{r}\right]\cdot\mathbf{P}=\frac{1}{2}\mathbf{B}(R)\cdot\left[\mathbf{r}\times\mathbf{P}\right],
\end{equation}
therefore, the total Hamiltonian \eqref{eq:total_H} reduces to
\begin{flalign}
    H=\frac{\mathbf{p}^2}{2\mu}+\frac{\mathbf{P}^2}{2M}+\frac{e^2B_z^2(R)r^2}{8\mu}+\frac{e}{M}\mathbf{B}\cdot\mathbf{r}\times\mathbf{P}+V(r). \label{eq:total_H2}
\end{flalign}

\subsection{Relative motion}\label{sub.relative}

Now, we consider the Born-Oppenheimer approximation, as the electron-hole relative motion occurs at a different scale (faster) than the center of mass motion, and solve first the problem of the relative motion ignoring the kinetic energy of the center of mass. However, note that the Hamiltonian \eqref{eq:total_H2} does not include the interaction between the magnetic field and the spin. Owing to incorporate that, knowing of the crucial relevant valley-spin locking physics in TMDs \cite{braganca2019magnetic, sallen2012robust, zeng2012valley, mak2012control, cao2012valley, xiao2012coupled, rasmita2021opto, Stier2016}, one adds 
\begin{equation}
    H_Z= \frac{ge}{4\hbar m_0}  \mathbf{B}(R)\cdot \mathbf{S},
\end{equation}
into Eq.~\eqref{eq:total_H2}, where $\mathbf{S}$ is the total spin of the electron-hole pair that lifts the valley degeneracy in the monolayer TMD energy spectrum. It corresponds to a Zeeman term where $g$ is the intervalley Landé g-factor, i.e., for a constant and homogeneous magnetic field, the splitting of the energies between excitons in different valleys becomes $g\mu_B B$, with $\mu_B=e\hbar/m_0$ being the Bohr's magneton \cite{Wozniak2020}. Early experimental studies of the magnetic field dependence of excitons in monolayer MoSe$_2$ obtained a Zeeman shift of $g\approx 4e/m_0$; this is related to the d-orbital character of the conduction and valence states around the $K$ and $K^\prime$ valleys of the TMD energy spectrum and their non-null orbital magnetic moments. \cite{li2014valley, macneill2015breaking} 

Therefore, for the relative motion, neglecting the center-of-mass momentum, the Hamiltonian to be considered, which depends on $\mathbf{R}$ and $\mathbf{P}$ as parameters, is
\begin{align}
    H_r&=\frac{\mathbf{p}^2}{2\mu}+\frac{e^2B_z^2(R)r^2}{8\mu}+\frac{e}{M}\mathbf{B}(R)\cdot\mathbf{r}\times\mathbf{P}\nonumber\\ 
    & \hspace{0.5cm} + \frac{ge}{4\hbar m_0} \mathbf{B}(R)\cdot \mathbf{S} + V(r). \label{eq:Hr}
\end{align}
Notice that Eq.~\eqref{eq:Hr} is composed of the kinetic energy of the relative motion, the interaction potential $V(r)$, two linear terms in the magnetic field, one of them that depends on the gradient $\nabla_\mathbf{R}$ and the other one on the total spin $\mathbf{S}$, and a quadratic contribution on the field. In view of that, one can consider perturbation theory in the magnetic field to solve it, such as $H_r=H_0+H_I$, adopting the unperturbed Hamiltonian $H_0$ as
\begin{equation}
    H_0=\frac{\mathbf{p}^2}{2\mu}+V(r), \label{eq:unperturbed}
\end{equation}
whose solutions to the eigenvalue problem
\begin{equation}
H_0\psi_{n,m,\tau}=E_{n,m,\tau}\psi_{n,m,\tau}, \label{eq:H0}
\end{equation}
can be obtained numerically, with $n$ and $m$ being the principal and angular quantum numbers, respectively, and $\tau$ being the valley-spin index. Here, we shall restrict ourselves to the $A$-excitons, i.e., to the quantum number case with $n=1$ and $m=0$, corresponding to electron and hole forming a bright exciton coming from a two-band system in the effective mass regime with the same spin and valley indexes [see Fig.~\ref{Fig1}(b)].

\begin{figure}[!t]
\includegraphics[width=\linewidth]{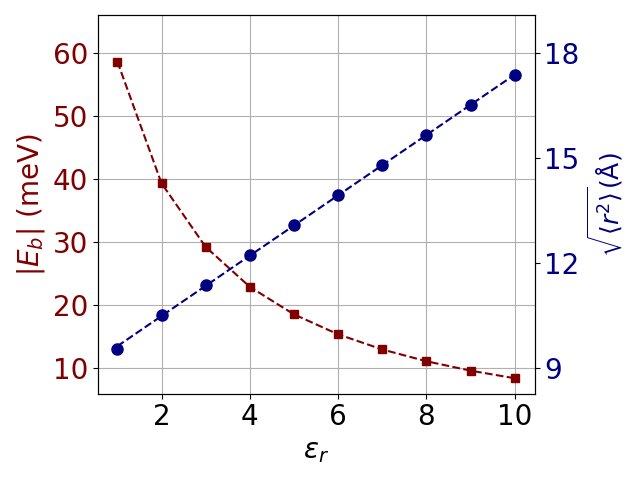}
\caption{\textcolor{blue}{(Color online)} Dependence of the exciton binding energy ($|E_b|$ - squared red symbols) and average radius of the exciton wavefunction ($lu\langle r^2 \rangle$ - circular blue symbols) with the environment dielectric constant $\epsilon_r$ for MoS$_2$.}
\label{Fig3}
\end{figure}

We choose to solve Eq.~\eqref{eq:unperturbed} in momentum space, where the Rytova-Keldysh potential is given by \cite{van2018coulomb, goryca2019revealing, tenorio2023tunable}
\begin{equation}\label{eq.V.RK}
V(q)=-\frac{e^2}{2\epsilon q}\frac{1}{\epsilon_r+r_0q},
\end{equation}
where $\epsilon_r$ is the relative dielectric constant and $r_0$ the material screening length. By Fourier transforming the Schr\"odinger equation taking the $s-$exciton, one gets the following integral expression
\begin{equation}\label{eq.integral.equation}
    u(q)=\int \frac{d\mathbf{q^\prime}}{(2\pi)^2} V(|\mathbf{q}-\mathbf{q^\prime}|) \frac{u(q^\prime)}{E+\frac{\hbar^2 {q^\prime}^2}{2\mu}},
\end{equation}
with $u(q)=\left[E+\hbar^2 q^2/(2\mu)\right]\psi(q)$. Here, for solving the integral equation \eqref{eq.integral.equation}, we use a traditional Gauss-Legendre quadrature method to discretize the integral, resulting in a problem that is solved by searching the eigenvalues $E<0$ that satisfy the linear system. The assumed effective mass value was obtained from Ref.~\cite{Kormanyos_2015}, that implies in a reduced mass of $\mu=0.25m_0$, and the $r_0$ parameter was extracted from Ref.~\cite{Berkelbach2013}, that in the case of monolayer MoS$_2$ it is $r_0=41.5$ \AA. By solving Eq.~\eqref{eq.integral.equation}, one finds the binding energy and the exciton wavefunction corresponding to the exciton relative motion for a fixed surrounding system environment, i.e., a fixed $\epsilon_r$. Thus, we show in Fig.~\ref{Fig3} how the binding energy and average radius of the exciton relative motion depend on the dielectric constant $\epsilon_r$. As expected by Eq.~\eqref{eq.V.RK}, a large dielectric constant screens more the electron-hole interaction, and consequently, it lowers the exciton binding energy, as observed by the dashed brown curve in Fig.~\ref{Fig3}. Due to this reduction in the interaction strength $V(r)$, the exciton wavefunction will be more spread, as can be noticed by the monotonic growth of the average radius of the exciton wavefunction shown by the dashed blue curve in Fig.~\ref{Fig3}.

For the perturbative term $H_I$ that is left from Eq.~\eqref{eq:Hr}, i.e.
\begin{equation}
    H_I=\frac{e^2B_z^2(R)r^2}{8\mu}+\frac{e}{M}\mathbf{B}(R)\cdot\mathbf{r}\times\mathbf{P}+ \frac{ge}{4\hbar m_0} \mathbf{B}(R)\cdot \mathbf{S}, \label{eq:HI}    
\end{equation}
one needs to calculate its expectation value. Note that the $\mathbf{B}(R)\cdot\mathbf{S}$ term is the one responsible for breaking the valley degeneracy in the system, as shall appear explicitly by a valley-spin index $\tau$ in its expectation value. At the lowest order of perturbation theory, the expectation value of Eq.~\eqref{eq:HI} is computed with the solution of Eq.~\eqref{eq:H0}, resulting in
\begin{equation}
E_\tau(R)=\frac{e^2B_z^2(R)a_0^2}{8\mu} + \tau g \frac{\hbar}{2}B_z(R), \label{eq:ER}
\end{equation}
with $\tau=+1$ for a spin/valley combination and $-1$ for the other, $a_0=\sqrt{\langle r^2\rangle}$ the exciton Bohr radius. We neglected the contributions that appear in second-order perturbation theory and would imply an effective mass dependence on the radial coordinate $R$, such as a renormalization of the exciton mass, due to the second term in Eq.~\eqref{eq:HI} that is proportional to $\mathbf{P}$ \cite{Freire2000}. The eigenvalue of the relative motion problem, i.e., the sum of the eigenvalue of the unperturbed problem of Eq.~\eqref{eq:H0} plus the eigenvalue of the first-order perturbation problem given by Eq.~\eqref{eq:ER}, corresponds to the effective potential that will trap the exciton center of mass. For the investigated case here of bright s-excitons for monolayer TMDs, it corresponds to 
\begin{equation}\label{eq.Veff}
    V_{eff}^\tau(R)= E_{1,0,\tau} + E_\tau(R).
\end{equation} 
Notice that $E_\tau(R)$ in Eq.~\eqref{eq:ER} is composed of linear and quadratic contributions on the magnetic field coming from the Zeemann and diamagnetic terms, respectively. These contributions will be of the same magnitude when
\begin{equation}
B_z(R)\sim\frac{4g\hbar\mu}{e^2a_0^2}, \label{eq:BzC}
\end{equation}
that for monolayer TMDs case, using typical values of $g$ in the range of $1.57 e/m_0$ to $4.8 e/m_0$ \cite{Wozniak2020}, taking an exciton radius of the order of $1$ nm \cite{Gang2018}, and the reduced mass of about $0.25m_0$ \cite{Kormanyos_2015}, it implies that only magnetic fields of the order of $660$ T make both the diamagnetic and Zeemann terms to be the same order of magnitude.

\subsection{Center of mass motion}\label{sub.center}

\begin{figure}[!t]
\includegraphics[width=0.9\linewidth]{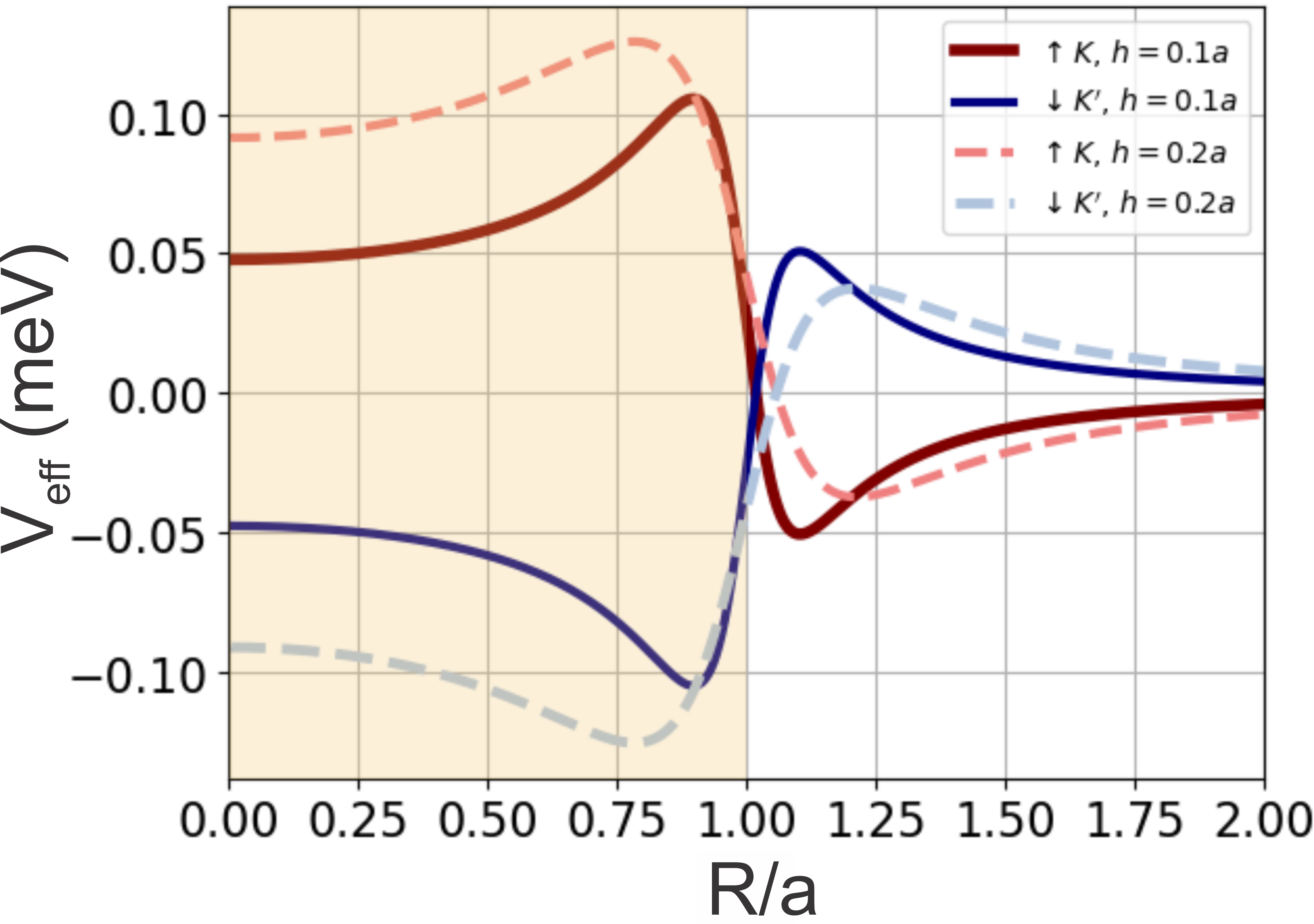}
\caption{\textcolor{blue}{(Color online)} Effective potential [Eq.~\eqref{eq.Veff}] for the exciton center of mass due to both the diamagnetic and Zeeman terms for both $\tau=+1$ ($\uparrow$ red curves) and $\tau=-1$ ($\downarrow$ blue curves) valleys, taking the following fixed parameters: $\mu_0{\cal M}=1.33$ T, $d_{hBN}=1.0$ nm, $a=1.0~\mu$m. Solid (dashed) curves correspond to the case for $h=0.1a$ ($h=0.2a$).}
\label{Fig4}
\end{figure}

After solving the problem for the relative motion, we now focus on the center of mass motion whose Hamiltonian $H_{CM}$ is given by the sum of the center of mass kinetic energy and the effective potential given by Eq.~\eqref{eq.Veff}. The latter is seen as the exciton trap potential of the center of mass motion, composed by adding the expectation value \eqref{eq:ER} of the first-order perturbative Hamiltonian $H_I$ and the eigenvalue of the unperturbed solution \eqref{eq:H0} in the center of mass Hamiltonian. It reads as
\begin{flalign}
    H_\mathrm{CM}=\frac{\mathbf{P}^2}{2M}+V_{eff}^\tau(R). \label{eq:Hcom}
\end{flalign}
In Fig.~\ref{Fig4}, we show the effective potential $V_{eff}^\tau(R)$ for different magnetic disk thickness $h$, fixing the distance between the disk and the TMD layer as $z=h+d_{hBN}$ with $d_{hBN}=1.0$ nm (corresponding to a slab of three hBN monolayers), the disk radius of $a=1.0~\mu$m, and a magnetization of $\mu_0 {\cal M}=1.33$ T for different valleys. One notices that for one of the valleys, the potential minimum is at $R/a<1$ (yellow shaded region) and, therefore, the exciton will be confined near the border inside the disk (blue curves for $\downarrow$). In contrast, for the other valley, the exciton will be confined outside the disk (red curves for $\uparrow$) since the potential minimum is at $R/a>1$. Moreover, when one increases the magnetic disk thickness from $h=0.1a$ (solid curves) to $h=0.2a$ (dashed curves), the inner potential becomes deeper while also becoming wider, whereas the outer potential becomes shallower. From this analysis of the effective potential [Fig.~\eqref{Fig4}], we can already expect a valley-dependent spatial separation of the excitons.

\begin{figure*}[!t]
    \centering
      \includegraphics[width=0.9\linewidth]{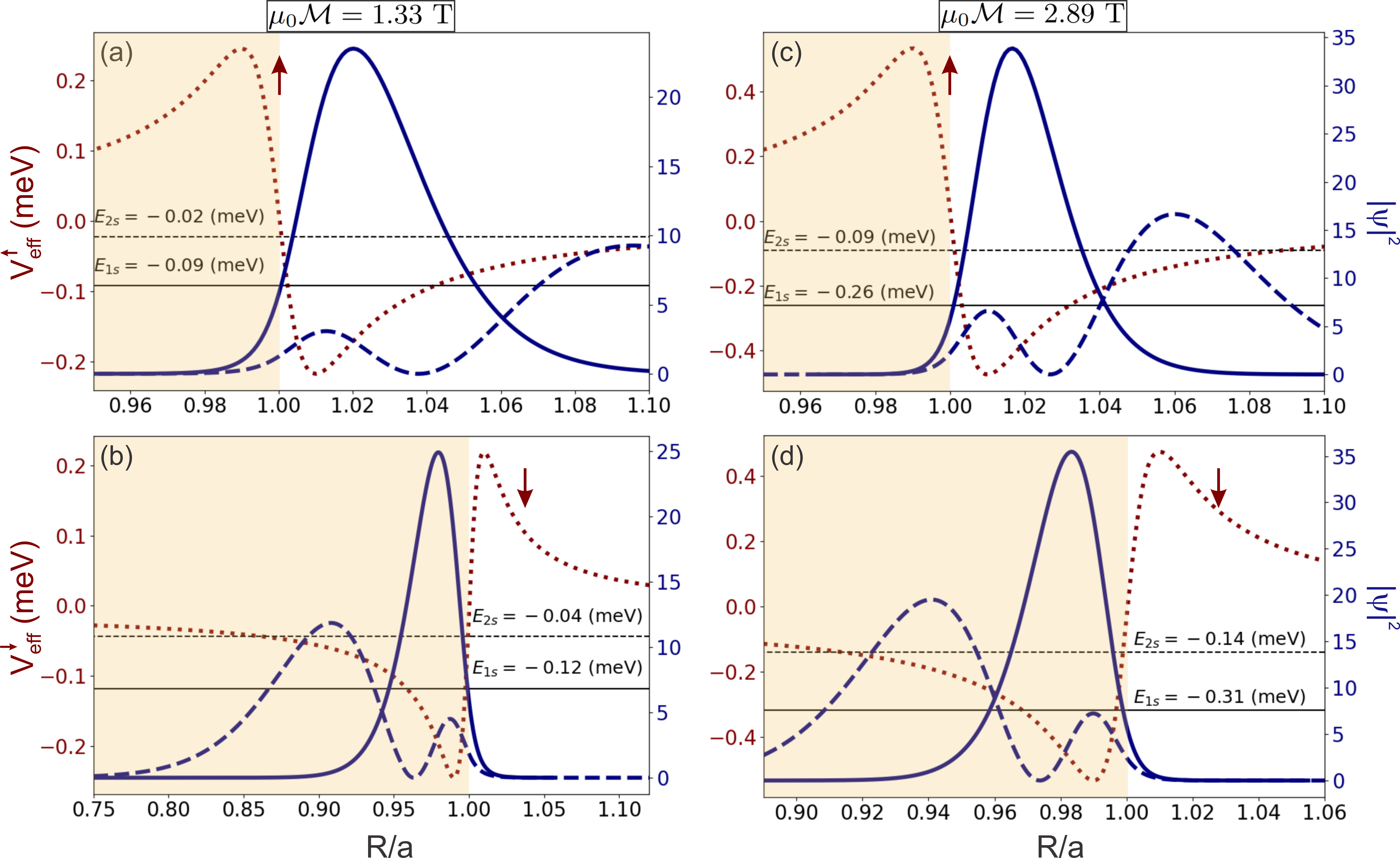}
      \caption{\textcolor{blue}{(Color online)} Eigenfunctions (blue curves) and eigenvalues (black curves) of the first (solid curves) and second (dashed curves) excitonic states of the center of mass problem given by Eq.~\eqref{eq:Hcom} for different system magnetization and valley-spin index. Top panels [(a), (c)] and bottom panels [(b), (d)] corresponds to $\uparrow$ ($\tau=+1$ -- K) and $\downarrow$ ($\tau=-1$ -- K$^\prime$) valley-spin index, respectively. Left (right) panels are for the magnetization $\mu_0{\cal M}=1.33$ T ($\mu_0{\cal M}=2.89$ T). It was taken the following system parameters: $a=1.0~\mu$m, $h=0.1a$, and $d_{hBN}=1.0$ nm. Red dashed curves are the effective potential [Eq.~\eqref{eq.Veff}] related to the left plot axis. Right axis is relative to the wavefunction plots.}
      \label{fig5}
\end{figure*}

We calculate the wavefunctions and eigenvalues of Eq.~\eqref{eq:Hcom} by the shooting method, solving the corresponding radial differential equation with an explicit five-order Runge-Kutta method \cite{Dormand1980} as implemented in the script package of Ref.~\cite{2020SciPy-NMeth}. The results are depicted in Fig.~\ref{fig5} for different magnetization: (left panels) $\mu_0{\cal M}=1.33$ T and (right panels) $\mu_0{\cal M}=2.89$ T; and different spin-valley index: (top panels) $\uparrow$ and (bottom panels) $\downarrow$. Figures~\ref{fig5}(a) and \ref{fig5}(c) for $\tau=+1$ show the first two exciton states (1s and 2s) confined outside the disk, while in Figs.~\ref{fig5}(b) and \ref{fig5}(d) one observes the exciton states confined inside the disk. Regarding the nature of the confined exciton states at the disk border, one can realize that the exciton binding energies for the $\downarrow$--states confined inside the disk are energetically lower in magnitude than the $\uparrow$--states confined outside the disk, i.e., the $\downarrow$--states are slightly more bounded than the $\uparrow$--states and, consequently, the exciton wavefunctions of the $\uparrow$--states are more spread out when compared with the excitons inside the disk, denoted by the shaded yellow region. It is interesting to note that the wavefunction peaks of the first exciton states confined inside and outside of the disk for valley-spin index $\tau=-1$ and $\tau=+1$, respectively, are spatially separated by a distance of about $0.04a$, which corresponds to a large spatial valley-spin resolved exciton separation of about $40$ nm, making it possible to be measured experimentally for feasible system parameters. Moreover, one observes that the exciton wavefunctions are asymmetrical due to the asymmetry in the potential profile, which is sharp in the direction of the disk border but smooth when distancing from the border; this explains the asymmetry in the double peak structure of the second exciton state (2s). By comparing the left and right panels for disk-induced magnetization of $\mu_0{\cal M}=1.33$ T and $\mu_0{\cal M}=2.89$ T, respectively, one obtains that the exciton energies are more energetically bounded and consequently, the wavefunctions are less spread out the higher the magnetization amplitude, as expected since the potential becomes deeper as the magnetization increases.

\section{Discussions and Final remarks}\label{sec.discussions}

In summary, we have theoretically investigated exciton confinement in monolayer TMDs induced by a magnetized disk placed on the top of an hBN slab that covers the TMD, as illustrated in Fig.~\ref{Fig1}. The magnetized disk generates an inhomogeneous magnetic field, whose abrupt character implies an intense magnetic field at its border. Due to the Zeeman term, we showed that the exciton center of mass motion is subjected to an effective potential obtained from the relative motion solution, exhibiting a minimum located inside or outside the disk, that can confine excitons depending on the valley-spin state. As a consequence of the spin-orbit coupling in TMDs, this implies that the Zeeman term will act differently on each valley and will confine excitons inside or outside the magnetized disk, as shown in Fig.~\ref{fig5}, with a spatial separation of dozens of nanometers. The theoretical framework presented here can be applied to different van der Waals materials within the effective mass approximation as well as to different shapes of the magnetized structure placed on the top of the setup to induce the inhomogeneous magnetic field and create the effective potential for the exciton confinement.

Regarding the exciton energy scale, the binding energy for the center of mass is of the order of hundreds of $\mu$ eV, which is small when compared with the relative motion binding energy, which was obtained of the order of hundreds of meV. With respect to the magnetized structure and the exciton wave function, we have shown results for the disk case, being the exciton wavefunction confined very close to the disk border. Based on that, it is to be expected similar results by considering instead of a disk, a stripe, or an array of stripes as discussed in Refs.~\cite{Nogaret_2010, nogaret1997observation} for conventional semiconductors. For the strip setup, we can expect that the valley excitons will confine in one of the directions while the exciton is free to move along the growth direction of the stripe.

Knowing that the Zeeman term is the one responsible for breaking the valley degeneracy and, consequently, leading the valley-spin resolved exciton localization in our problem, it is worth commenting on the role of the $g$-factor on the Zeeman term and how it can enhance the exciton confinement effect. It has been reported in the literature high $g$-factors in van der Waals structures \cite{Forste2020} and twist-dependent $g$-factor in twisted hetero-bilayer structures \cite{seyler2019signatures}. Thus, high $g$-factor structures will greatly enhance the exciton confinement due to the Zeeman term. Also, for the case of a superconducting disk, we can expect magnetic fields of one order of magnitude higher than the case of the magnetized disk \cite{Freire2000}; thus, we can infer that a system of a hetero-bilayer on top of a superconducting disk is the ideal situation for the trapping of valley excitons at different spatial locations. In addition, we emphasize that the diamagnetic term for the case studied here is totally negligible. Still, from Eq.~\eqref{eq:BzC}, we can infer that in a system with low reduced mass, low g-factor, and high exciton Bohr radius, the diamagnetic term can be at the same order of magnitude as the Zeeman term; however, the former does not break the valley degeneracy.

Concerning the viability of experimental measuring the spatial and valley-resolved exciton confinement effect observed here, we emphasize that: (i) the binding energy that we reported is of the order of 1 K. Such low temperatures can be obtained with dilution refrigerators for exciton measurements, as depicted in Refs.~\cite{ugeda2014giant, zhang2014direct, ross2013electrical, he2014tightly, chen2019luminescent}; (ii) the valley excitons can be probed with different circularly polarized light \cite{cao2012valley, mak2012control, zeng2012valley, sallen2012robust}; thus, it is expected that a photo-luminescence experiment, with a resolution of few nanometers, would be able to detect the circular dichroism. 

\section*{Acknowledgements}

N.M.R.P. acknowledges support by the Portuguese Foundation for Science and Technology (FCT) in the framework of the projects PTDC/FIS-MAC/2045/2021 and EXPL/FISMAC/0953/2021, and the Strategic Funding UIDB/04650/2020. N.M.R.P. also acknowledges the Independent Research Fund Denmark (grant no. 2032-00045B) and the Danish National Research Foundation (Project No. DNRF165). D.R.C and A.J.C. were supported by the Brazilian Council for Research (CNPq) through Universal and PQ programs, and the Brazilian National Council for the Improvement of Higher Education (CAPES). D.R.C gratefully acknowledges the support from CNPq grants $313211/2021-3$, $437067/2018-1$, $423423/2021-5$, $408144/2022-0$ and the Research Foundation—Flanders (FWO).

\appendix
\section{Analytical expression for the magnetic field} \label{app:Bz}

Within the $h\rightarrow 0$ limit, the magnetic field $B_z$ \eqref{eq:Bfield} can be obtained by taking the derivative of Eq.~\eqref{eq:Afield_c} of the vector potential, resulting in
\begin{widetext}
\begin{eqnarray}\label{eq:Bfield.final}
B(r,z)=\frac{1}{2r}A(r,z)
+\frac{2\mu_0h{\cal M}\sqrt{ar}}{\pi k r^2}\left[1+\frac{k^2}{2}\left(1-\frac{r}{a}\right) \right]
 \left[\frac{1}{\sqrt{1+k^2}}E\left(\frac{k^2}{k^2+1} \right)
- K(-k^2)+ \frac{1}{k^2+1}E(-k^2)
 \right].
 \end{eqnarray}\end{widetext}

\bibliography{references}

\end{document}